\documentclass[twocolumn]{aastex63}
\usepackage{comment}
\usepackage{color}
\received{}
\revised{}
\accepted{\today}
\graphicspath{{./}{figures/}}

\begin{document}

\title{
The Core Mass Function in the Orion Nebula Cluster Region:
What Determines the Final Stellar Masses?}

\author{Hideaki Takemura}
\affil{The Graduate University for Advanced Studies
(SOKENDAI), 2-21-1 Osawa, Mitaka, Tokyo 181-0015, Japan}
\affil{National Astronomical Observatory of Japan, 2-21-1 Osawa, Mitaka, Tokyo 181-8588, Japan}

\author{Fumitaka Nakamura}
\affil{The Graduate University for Advanced Studies
(SOKENDAI), 2-21-1 Osawa, Mitaka, Tokyo 181-0015, Japan}
\affil{National Astronomical Observatory of Japan, 2-21-1 Osawa, Mitaka, Tokyo 181-8588, Japan}
\affil{Department of Astronomy, The University of Tokyo, Hongo, Tokyo 113-0033, Japan}

\author{Shuo Kong}
\affiliation{Steward Observatory, University of Arizona, Tucson, AZ 85719, USA}
\affil{Department of Astronomy, Yale University, New Haven, CT 06511, USA}

\author{Héctor G. Arce}
\affil{Department of Astronomy, Yale University, New Haven, CT 06511, USA}

\author{John M. Carpenter}
\affil{Joint ALMA Observatory, Alonso de Córdova 3107 Vitacura, Santiago, Chile}

\author{Volker Ossenkopf-Okada}
\affiliation{I.~Physikalisches Institut, Universit\"at zu K\"oln,
              Z\"ulpicher Str. 77, D-50937 K\"oln, Germany}

\author{Ralf Klessen}
\affiliation{Universit\"{a}t Heidelberg, Zentrum f\"{u}r Astronomie, Albert-Ueberle-Str. 2, 69120 Heidelberg, Germany}
\affiliation{Universit\"{a}t Heidelberg, Interdisziplin\"{a}res Zentrum f\"{u}r Wissenschaftliches Rechnen, INF 205, 69120 Heidelberg, Germany}

\author{Patricio Sanhueza}
\affil{The Graduate University for Advanced Studies
(SOKENDAI), 2-21-1 Osawa, Mitaka, Tokyo 181-0015, Japan}
\affil{National Astronomical Observatory of Japan, 2-21-1 Osawa, Mitaka, Tokyo 181-8588, Japan}

\author{Yoshito Shimajiri}
\affil{National Astronomical Observatory of Japan, 2-21-1 Osawa, Mitaka, Tokyo 181-8588, Japan}

\author{Takashi Tsukagoshi}
\affil{National Astronomical Observatory of Japan, 2-21-1 Osawa, Mitaka, Tokyo 181-8588, Japan}

\author{Ryohei Kawabe}
\affil{The Graduate University for Advanced Studies
(SOKENDAI), 2-21-1 Osawa, Mitaka, Tokyo 181-0015, Japan}
\affil{National Astronomical Observatory of Japan, 2-21-1 Osawa, Mitaka, Tokyo 181-8588, Japan}

\author{Shun Ishii}
\affil{National Astronomical Observatory of Japan, 2-21-1 Osawa, Mitaka, Tokyo 181-8588, Japan}

\author{Kazuhito Dobashi}
\affil{Tokyo Gakugei University, Koganei, Tokyo, 184-8501, Japan}

\author{Tomomi Shimoikura}
\affil{Otsuma Women’s University, ‘Chiyoda-ku, Tokyo, 102-8357, Japan}

\author{Paul F. Goldsmith}
\affil{Jet Propulsion Laboratory, California Institute of Technology, 4800 Oak Grove Drive, Pasadena, CA 91109, USA}

\author{\'Alvaro S\'anchez-Monge}
\affiliation{I.~Physikalisches Institut, Universit\"at zu K\"oln,
              Z\"ulpicher Str. 77, D-50937 K\"oln, Germany}

\author{Jens Kauffmann}
\affiliation{Haystack Observatory, Massachusetts Institute of Technology, 99 Millstone Road, Westford, MA 01886, USA}

\author{Thushara Pillai}
\affiliation{Max--Planck--Institut f\"ur Radioastronomie, Auf dem H\"ugel 69, D--53121 Bonn, Germany}

\author{Paolo Padoan}
\affil{Institut de Ci\' encies del Cosmos, Universitat de Barcelona, IEEC-UB, Mart\'i i Franqu\'es 1, E08028 Barcelona, Spain}
\affil{ICREA, Pg. Llu\'is Companys 23, E-08010 Barcelona, Spain}

\author{Adam Ginsberg}
\affil{Department of Astronomy, University of Florida, Gainesville, FL 32611, USA}

\author{Rowan J. Smith}
\affiliation{Jodrell Bank Centre for Astrophysics, School of Physics and Astronomy, University of Manchester, Oxford Road, Manchester M13 9PL, UK}

\author{John Bally}
\affiliation{Department of Astrophysical and Planetary Sciences, University of Colorado, Boulder, Colorado, USA}

\author{Steve Mairs}
\affiliation{East Asian Observatory, 660 N. A`oh\={o}k\={u} Place, Hilo, Hawai`i, 96720, USA}

\author{Jaime E. Pineda}
\affil{Max-Planck-Institut f\" ur extraterrestrische Physik, Giessenbachstrasse 1, D-85748 Garching, Germany}

\author{Dariusz C. Lis}
\affiliation{Jet Propulsion Laboratory, California Institute of Technology, 4800 Oak Grove Drive, Pasadena, CA 91109, USA}

\author{Blakesley Burkhart}
\affil{Department of Physics and Astronomy, Rutgers, The State University of New Jersey, 136 Frelinghuysen Rd, Piscataway, NJ 08854, USA}


\author{Peter Schilke}
\affiliation{I.~Physikalisches Institut, Universit\"at zu K\"oln,
              Z\"ulpicher Str. 77, D-50937 K\"oln, Germany}

\author{Hope How-Huan Chen}
\affil{Department of Astronomy, The University of Texas, Austin, TX 78712, USA}

\author{Andrea Isella}
\affil{Department of Physics and Astronomy, Rice University, 6100 Main Street, Houston, TX 77005, USA}

\author{Rachel K. Friesen}
\affil{National Radio Astronomy Observatory, 520 Edgemont Rd., Charlottesville, VA, 22903, USA}

\author{Alyssa A. Goodman}
\affil{Harvard-Smithsonian Center for Astrophysics, 60 Garden Street, MS 42, Cambridge, MA 02138, USA}

\author{Doyal A. Harper}
\affil{Department of Astronomy and Astrophysics University of Chicago, Chicago, IL 60637, USA}

\begin{abstract}
Applying dendrogram analysis to the CARMA-NRO C$^{18}$O ($J$=1--0) data
having an angular resolution of $\sim$ 8$\arcsec$, we identified 692 dense cores in the Orion Nebula Cluster (ONC) region.
Using this core sample, we compare the core and initial stellar mass functions in the same area to quantify the step from cores to stars.
About 22 \% of the identified cores are gravitationally bound.
The derived core mass function (CMF) for starless cores has a slope similar to Salpeter's stellar initial mass function (IMF) for the mass range above 1 $M_\odot$, consistent with previous studies. Our CMF has a peak at a subsolar mass of $\sim$ 0.1 $M_\odot$, which is
comparable to the peak mass of the IMF derived in the same area.
We also find that the current star formation rate is consistent with the picture in which
stars are born only from self-gravitating starless cores.
However, the cores must gain additional gas from the surroundings to reproduce the current IMF (e.g., its slope and peak mass), because the core mass cannot be accreted onto the star with a 100\% efficiency. Thus, the mass accretion from the surroundings may play a crucial role in determining the final stellar masses of stars.
\end{abstract}


\section{Introduction}
\label{sec:intro}

Stars are believed to form in the dense
parts of molecular clouds, called dense cores \citep[e.g.,][]{shu87}. However, the evolution of such cores, particularly the process of star formation, is a matter of debate. There are two scenarios widely-discussed so far: competitive accretion \citep{bonnell06} and core-collapse models \citep[e.g.,][]{{shu87,mckee03}}.
In the competitive accretion scenario, stellar seeds, which are formed from the local
dense
parts of the core,
initially have low-mass of $\sim$1 $M_\odot$, and
gain additional mass from
the surroundings through the modified Bondi-Hoyle accretion.
The mass function of dense cores (CMF) of the stellar seeds is likely to be different in shape from the stellar initial mass function (IMF) at least at the early evolutionary phase.
\citep[e.g.,][]{zinnecker82, goodwin08}.
In the core-collapse model, 
final stellar masses are largely determined by the masses of the progenitor cores. Thus, a one-to-one correspondence between the core mass and stars formed is likely to be a natural outcome.
A hybrid model,
the clump-fed model,
has also been
proposed \citep{wang10}, where massive cores preferentially located at the bottom of the gravitational potential tend to gain additional mass through accretion of ambient gas.
The converging flow and global gravitational collapse scenarios have also been widely-discussed and attempt to explain observational characteristics of star-forming regions \citep{klessen98,klessen10,vazquez19,ballesteros20}.
Recent numerical simulations have pointed out the importance of mass accretion in the evolution of dense cores \citep{padoan14,pelkonen20}
. These different scenarios lead to different CMFs. Thus, the observed characteristics of CMFs  provide a key to constraining the star formation scenarios.

Many previous studies of CMFs toward nearby star-forming regions have revealed that the CMFs resemble the IMF \citep[e.g.,][]{motte98,alves07}. 
For example, \citet{alves07} identified dense cores in the Pipe Nebula based on near-infrared extinction observations,
and showed that the CMF in the Pipe Nebula has a similar slope to the IMF of the Orion Nebula Cluster (ONC) but its turnover mass is somewhat larger than that of the IMF
at $\sim$ 0.1 $M_\odot$.
They suggest that if 30$-$40 \% of the core mass goes into a star or stellar system forming inside,
the turnover mass of the resultant IMF from the Pipe Nebula CMF would coincide with that of the IMF in the ONC region.
However, very recently, \citet{motte18} reported a shallower CMF in the high mass star-forming region, W33, and suggested a possibility of a time-evolved CMF.
This evolution is further investigated in \citet{sanhueza19} and \citet{kong19} in infrared dark clouds (IRDCs), considering the studies of \citet{liu18} and \citet{cheng18}.
The effect of a time-evolved CMFs is also discussed in 
detail
in \citet{clark07} and \citet{dib10}.
It is worth noting that \citet{kroupa19} also discussed the variations of IMFs from region to region.

In this Letter, we compare the CMF and IMF in the ONC region, 
using a high-angular resolution C$^{18}$O ($J$=1--0) map \citep{kong18}.
The core catalog in the whole Orion A cloud will be presented in a forthcoming paper.
Our analysis presented below is the first direct comparison between the CMF and the IMF in the ONC region, in the mass range from $10^{-1}\ M_\odot$ to $10^2\ M_\odot$.

\section{Observations and data}
\label{sec:obs}

\subsection{C$^{18}$O (J=1--0) data}

We use the wide-field C$^{18}$O ($J$=1--0, 109.782182 GHz) data from the CARMA-NRO Orion survey, for which we obtained high-resolution $^{12}$CO, $^{13}$CO, and C$^{18}$O maps of Orion A, by combining the data taken with the CARMA interferometer and the 
NRO 45-m single-dish telescope.
See \citet{kong18} for more detail. The angular resolution of the maps is about 8\arcsec, corresponding to 3300 au at a distance of 414 pc \citep{menten07}
\footnote{Based on the Gaia data, \citet{grobschedl18} estimated the distance of 390 pc. However, we use 414 pc in this Letter. If we adopt the updated distance, the core masses tend to be about 10\% smaller.}.
The velocity resolution is $\sim$ 0.1 km s$^{-1}$.
The mean noise level of the C$^{18}$O map is 0.70 K ($\approx$ 1$\sigma$) in units of T$_{\rm MB}$.
Our map covers a 1 $\times$ 2 square degree area, containing OMC-1/2/3/4, L1641N, and V380 Ori.
In this Letter, we use a part of the map including the OMC-1 region and the ONC region.
The integrated intensity map of the region of interest is presented in Figure \ref{fig:obs_area} (a).

\subsection{H$_2$ column density data}

We use the \textit{Herschel}--\textit{Planck} H$_2$ column density map to calculate the core masses.
The map is constructed by \citet{kong18} based on
the 250 $\mu$m emission map with a 16\arcsec \ resolution
and the dust temperature map with a 36\arcsec \ resolution. 
We note that the angular resolution of the H$_2$ map is twice that of our C$^{18}$O ($J$=1--0) map.
We regridded the H$_2$ map to match the grids of the C$^{18}$O ($J$=1--0) data.


\subsection{Catalog of young stellar objects (YSOs)}
\label{subsec:YSO}

We use the catalog of young stars in the ONC region obtained by \citet{dario12}. Their catalog includes 1619 stars whose masses and ages are derived with the DM98 model \citep{dantonia98}. The spatial distribution of stars is presented in Figure \ref{fig:obs_area} (b).
We also used the catalog of 74 Class 0 and Class I protostars in the observed region from Herschel Orion Protostar Survey (HOPS) \citep{furlan16}.
The catalog covers Orion A and Orion B region with the luminosity range from 0.06 $L_\odot$ to 607 $L_\odot$
It is worth noting that the completeness of the HOPS catalog is only about 50 \% \citep{megeath06}. Therefore, we miss a significant number of true protostellar cores.
However, as shown below, the number of starless cores is much larger, and the incomplete 
identification of protostellar cores may not influence the shape of the CMF significantly.

\begin{figure}
 \begin{center}
 \includegraphics[scale=0.3,bb=0 0 1384 1384]{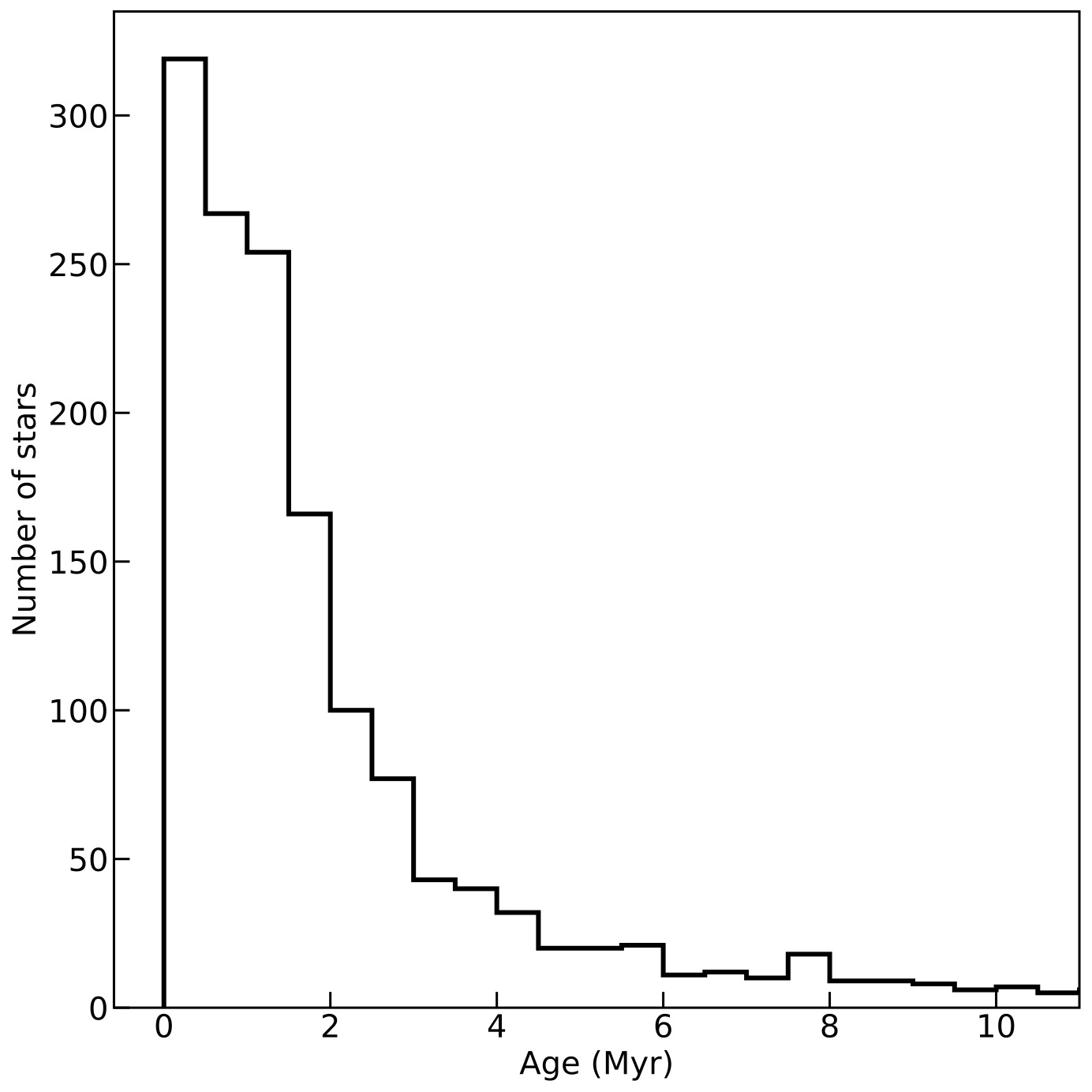}
 \end{center}
 \caption{The histogram of the stellar age in the ONC region from \citet{dario12} with the DM98 model.}
 \label{fig:age}
\end{figure}

Figure \ref{fig:age} shows the distribution of the young stars identified by \citet{dario12} as a function of age.
Most of the stars have inferred ages of about < 2 Myr with a tail to the age distribution out to ~ 10 Myr \citep[see][]{palla99}. However, as cautioned by \citet{hartmann01b}, the various observational uncertainties can create similar age distributions even if the underlying stellar population is coeval, which makes it difficult to robustly infer the star formation rate history. The estimated star formation rate is $\sim 1.5 \times 10^{-4}$ $M_\odot$ yr$^{-1}$ if the stars with inferred ages less than 2 Myr are considered.

\section{Dense cores and CMF in the ONC region}
\label{sec:result}

\subsection{Core Identification}
\label{sec:core_id}

\begin{figure}
 \begin{center}
 \includegraphics[scale=0.14,bb=0 0 3353 1478]{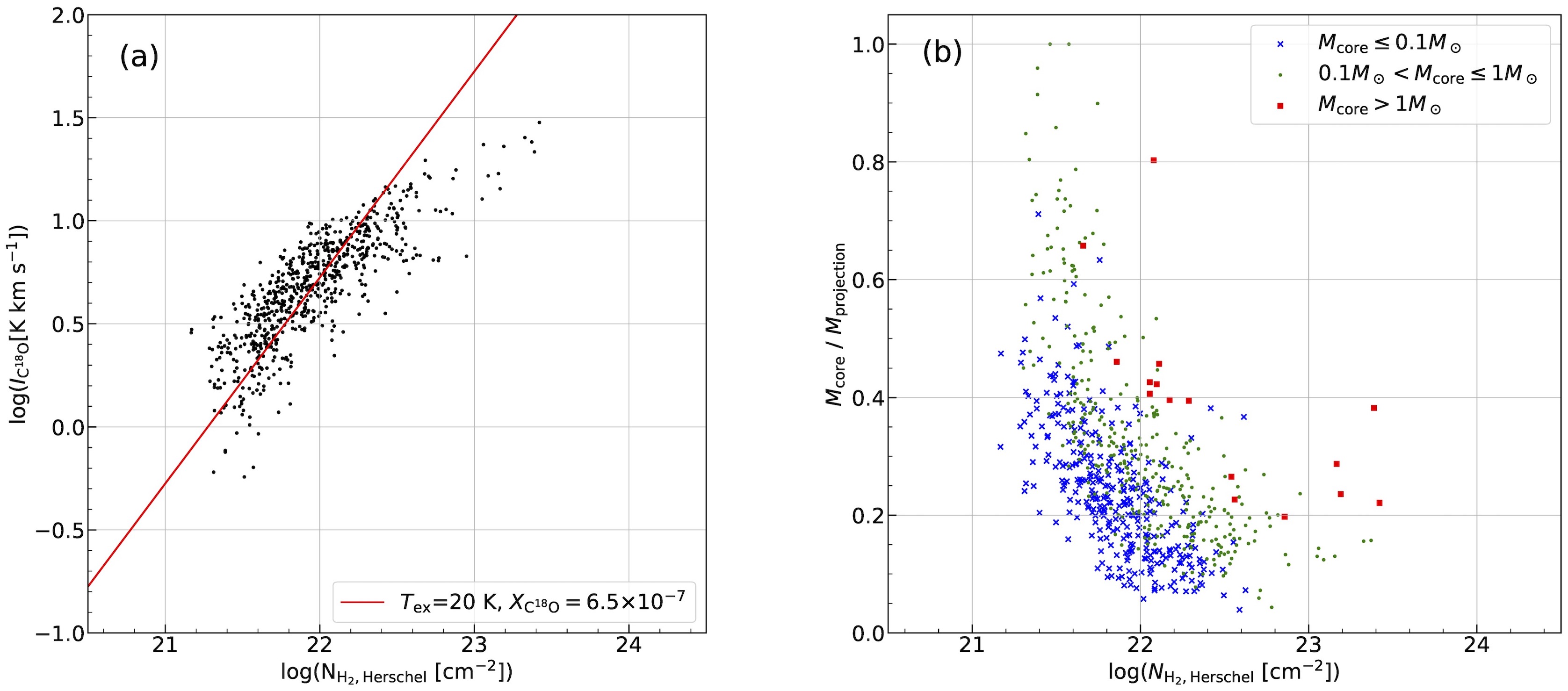}
 \end{center}
 \caption{(a) The correlation between the mean C$^{18}$O ($J$=1--0) integrated intensity
 and the mean \textit{Herschel}--\textit{Planck} H$_2$ column density in each projected core area.
 The solid line shows the best-fit function of optically thin LTE emission which is based on an
 abundance ratio of $X_\mathrm{C^{18}O}$=6.5$\times$10$^{-7}$ when $T_\mathrm{ex}$=20 K.
 (b) The relationship between the mass ratio, $M_\mathrm{core}/M_\mathrm{projection}$ and the \textit{Herschel}-- \textit{Planck} H$_2$ column density.}
 \label{fig:column_density}
\end{figure}

First, to verify whether our C$^{18}$O data can trace the dense structures reasonably well, we compare the C$^{18}$O column density with the H$_2$ column density derived from dust emission.
Figure \ref{fig:column_density} (a) indicates the correlation between the mean C$^{18}$O integrated intensity and the mean H$_2$ column density in each projected core area.
The solid line indicates the optically thin LTE emission. When $T_\mathrm{ex}$=20 K, the abundance ratio of C$^{18}$O with respect to H$_2$, $X_\mathrm{C^{18}O}$, is calculated as 6.5$\times$10$^{-7}$.
The C$^{18}$O integrated intensity is roughly proportional to the \textit{Herschel}--\textit{Planck} H$_2$ column density over the range of $\lesssim 10^{23}$ cm$^{-2}$.
Therefore, the C$^{18}$O emission is considered to be a reliable tracer of molecular hydrogen mass, and we use the C$^{18}$O emission to search for the dense structures in the molecular cloud.
However, there may be some effects of the CO depletion particularly in cold ($T \lesssim 20$ K), dense ($\gtrsim 10^5$ cm$^{-3}$) regions. In this sense, the total number of cores identified below may be somewhat underestimated.

We applied astrodendro ver. 0.2.0 \citep{rosolowsky08}\footnote{https://dendrograms.readthedocs.io/en/stable/} to the
C$^{18}$O ($J$=1--0) data cube to identify the cores by using the hierarchical structures of the molecular cloud.
Here, we define a {\it leaf} (the smallest structure identified by astrodendro) as a core. Then, we estimate the masses of the cores using the \textit{Herschel}--\textit{Planck} H$_2$ column density map, but we remove the contribution of the ambient gas distributed outside the cores in the position-position-velocity space.

From the CMF analysis with clumpfind, \citet{pineda09} pointed out that
the CMF shapes sometimes depend on the parameters of clumpfind, and recommended to use the core identification methods which take into account the cloud hierarchical nature, e.g., dendrogram.
Besides, recent synthetic observation studies applying dendrogram to the numerical simulation data showed that the structures identified in the PPP space are well related to the structures identified in the PPV space \citep{beaumont13, burkhart13}.
Thus, we believe that our definition of dense cores is reasonable for the statistical analysis of CMFs.

In the actual identification, the three input parameters of astrodendro are set to
min\_delta=1.4 K ($\approx$ 2$\sigma$), min\_value=1.4 K ($\approx$ 2$\sigma$), and min\_npix=60 ($\approx$ 1 beam $\times$ 3 channels), following the suggestions of \citet{rosolowsky08}.
Additional selection criteria are imposed to minimize the effect of the spatially varying noise levels for the core identification:
(1) the peak intensity of the {\it leaf} should be larger than $4\sigma$ at the corresponding spatial position,
(2) more than three successive channels should contain more than 20 pixels ($\approx $ a map angular resolution) for each channel.
In total, we identified 692 cores.

Then, we classify the cores into two groups, starless and protostellar cores, using the HOPS catalog.
If a core overlaps spatially with at least one HOPS object in the sky, we classified it as a protostellar core.
A core without overlapping HOPS objects is categorized as a starless core.
As a result, we identified 680 starless cores and 12 protostellar cores.
We note that almost all the HOPS class 0/I objects (20/21) are identified as {\it leaves}, but about half of such {\it leaves} are not satisfied with our additional condition (2). As a result, they are not classified into protostellar cores and we simply omit such cores in this Letter.

\begin{figure}
 \begin{center}
 \includegraphics[scale=0.22,bb=0 0 2249 978]{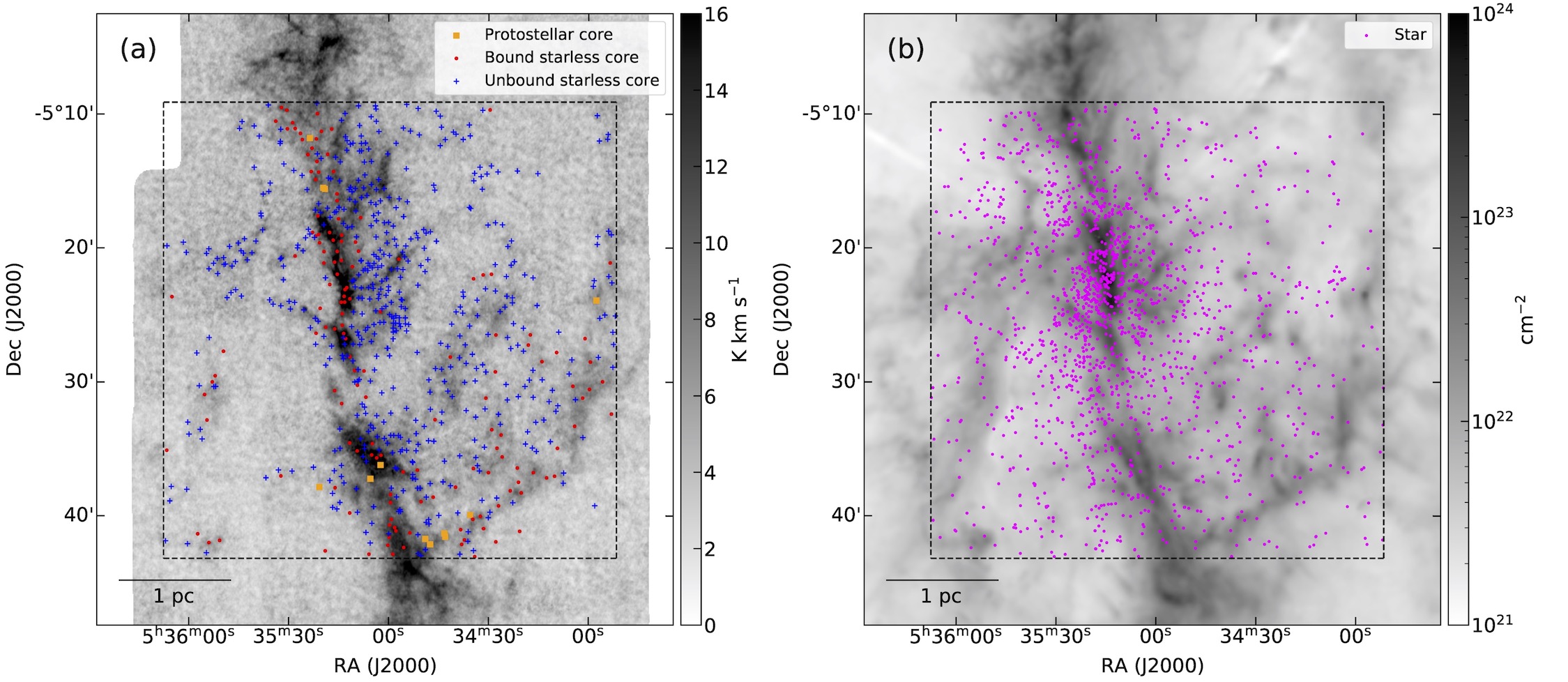}
 \end{center}
 \caption{(a) The C$^{18}$O integrated intensity map and (b) the \textit{Herschel}--\textit{Planck} H$_2$ column density map toward the ONC region as a dashed rectangle. The identified C$^{18}$O cores from this study and stars in the catalog of \citet{dario12} are plotted onto (a) and (b), respectively. In (a), the squares, circles, and crosses represent the protostellar cores, gravitationally bound starless cores, and unbound starless cores, respectively.}
 \label{fig:obs_area}
\end{figure}

\begin{figure}
 \begin{center}
 \includegraphics[scale=0.23,bb=0 0 2048 1536]{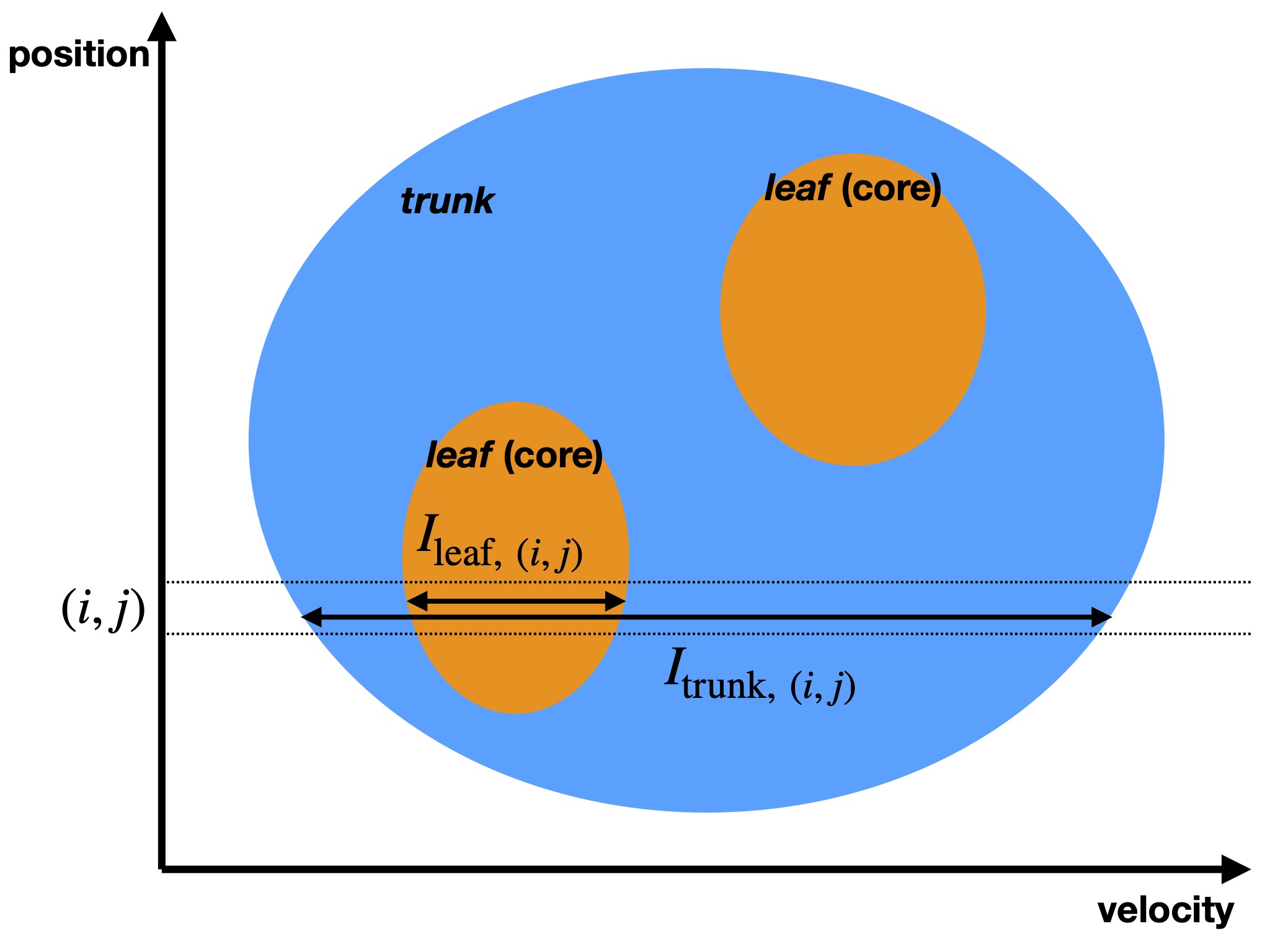}
 \end{center}
 \caption{Cartoon of dendrogram's hierarchical structures of intensity focusing on a {\it trunk} and a {\it leaf} in position-velocity plane. The two arrows represent the intensity of a {\it trunk} and a {\it leaf}.}
 \label{fig:intensity_ratio}
\end{figure}

Figure \ref{fig:obs_area} (a) shows the spatial distribution of starless and protostellar cores in the ONC region. The cores are distributed over the entire square box in Figure \ref{fig:obs_area} (a).
We calculated the core mass using the \textit{Herschel}--\textit{Planck} H$_2$ column density ($N_{\mathrm H_2}^{\rm Herschel}$) and intensity-ratio of the {\it leaf} and the {\it trunk} ($I_{\rm leaf}/I_{\rm trunk}$) (see Figure \ref{fig:intensity_ratio}).
We assigned the H$_2$ column density to each core using the intensity-ratio and calculated the core mass as 
$M_\mathrm{core}=3\times10^{-3}\times\sum (N_{\mathrm H_2}^{\rm Herschel} (i,\,j)/10^{22}\ \mathrm{cm^{-2}})\times I_{\mathrm{leaf}} (i,\,j)/I_{\mathrm{trunk}} (i,\,j)$, where $i$ and $j$ are the indices of the cell of interest on the R.A.--Dec. plane, respectively.
Figure \ref{fig:column_density} (b) shows the mass-ratio of $M_\mathrm{core}$, core mass, and $M_\mathrm{projection}$.
The mass $M_\mathrm{projection}$ is calculated by integrating all H$_2$ column density contained within the projection of cores (i.e., $M_\mathrm{projection}=3\times10^{-3}\times\sum (N_{\mathrm H_2}^{\rm Herschel} (i,\,j)/10^{22}\ \mathrm{cm^{-2}})$). The mean mass ratio is $\sim$ 0.29.

Using a virial analysis, we classify the starless cores into gravitationally-bound cores and unbound cores with the threshold of $\alpha_\mathrm{vir}$=$M_\mathrm{vir}/M_\mathrm{core}$=2 and also show their distribution in Figure \ref{fig:obs_area} (a).
For the virial mass $M_\mathrm{vir}$, we assumed a centrally condensed sphere without magnetic fields and external pressure as $M_\mathrm{vir}$=126($R_\mathrm{core}$/pc)($dV_\mathrm{core}$/km s$^{-1}$)$^2$.

We calculated the core radius $R_\mathrm{core}$ using a projected area of a core onto the plane of the sky, $A_\mathrm{core}$, as $R_\mathrm{core}=\sqrt{A_\mathrm{core}/\pi}$.
Here, the area, $A_\mathrm{core}$, and velocity dispersion, $v_\mathrm{rms}$ which is the intensity-weighted second moment of velocity, are calculated by astrodendro.
We categorized 151 starless cores, $\sim$22\% of starless cores, as gravitationally-bound starless cores.
We also calculated the core density as 
$\rho_\mathrm{core}$ as $\rho_\mathrm{core}={M_\mathrm{core}/(4/3)\pi R_\mathrm{core}^3}$.
The mean values and standard deviations of diameters, velocity widths in FWHM, masses, densities and virial ratios of the starless cores are $0.065\pm0.022$ pc, $0.34\pm0.13$ km s$^{-1}$, $0.19\pm0.42$ $M_\odot$, $(2.4\pm4.6)\times10^4$ cm$^{-3}$, $4.8\pm4.1$, respectively.

\section{CMFs in the ONC region}
\label{sec:discussion}
\begin{figure}
 \begin{center}
 \includegraphics[scale=0.33,scale=0.5,bb=0 0 2933 1315]{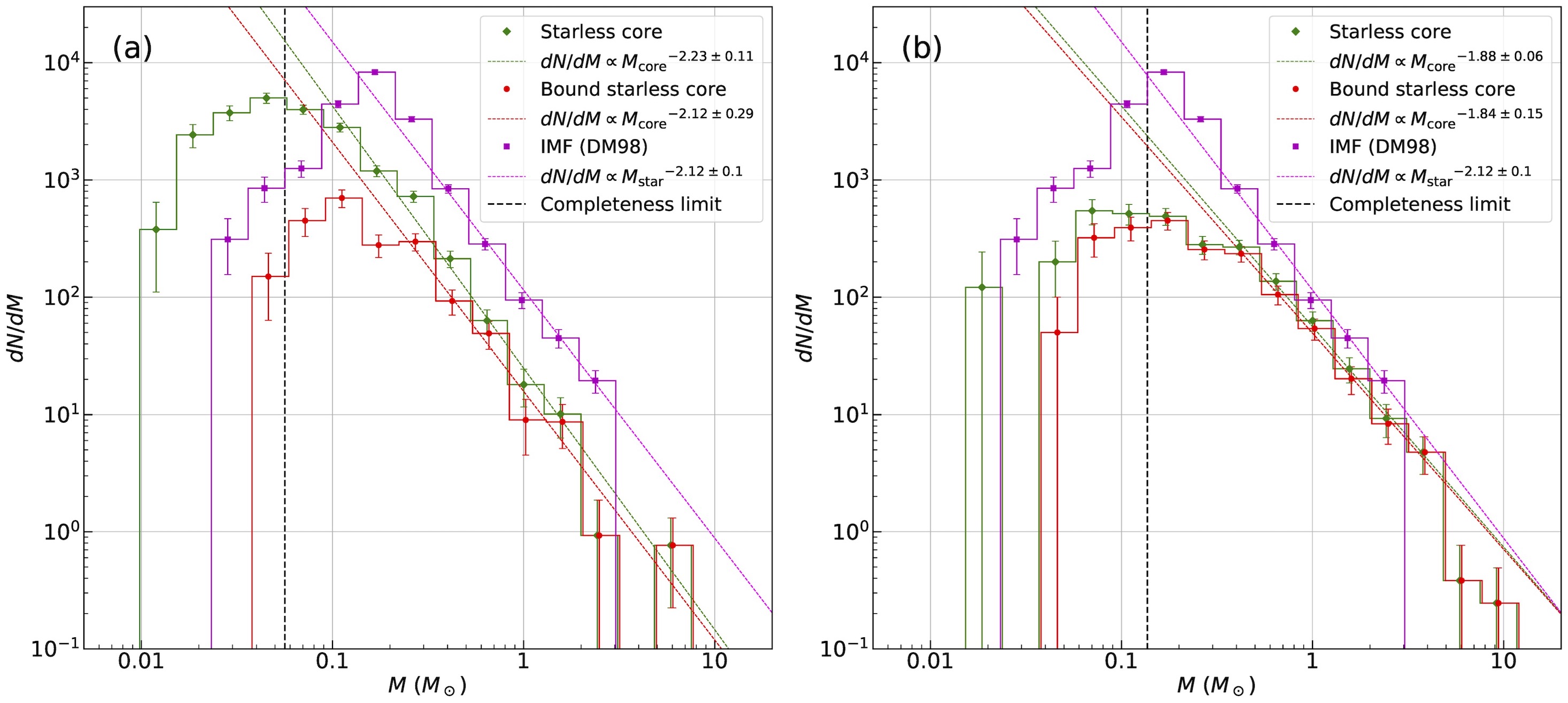}
 \end{center}
 \caption{CMFs in the ONC region.
 The astrodendro's parameters are(a) min\_delta=2$\sigma$, min\_value=2$\sigma$ and min\_npix=60
 and (b) min\_delta=3$\sigma$, min\_value=3$\sigma$ and min\_npix=120, respectively.
 For comparison, we show the IMF derived by \citet{dario12} in both panels.
 The 90 \% completeness limit is indicated with the vertical dashed lines.
 The error bars denote statistical errors.}
 \label{fig:cmf}
\end{figure}

Figure \ref{fig:cmf} (a) shows the CMFs toward ONC for all the starless cores and self-gravitating starless cores.
For comparison, we show the stellar IMF in Figure \ref{fig:cmf}.
The shapes of the CMFs are similar to those of the stellar IMF.
All CMFs have best-fit power-law indices of $\sim-$2 at the high-mass end.
The CMFs for all starless cores and self-gravitating cores have the turnover masses of $\sim$0.05 $M_\odot$ (below the completeness limit) and $\sim$0.11 $M_\odot$, respectively, which are comparable to that of the IMF.

The results of the core identification depend on the adopted parameters of astrodendro.
When we set min\_delta=3$\sigma$, min\_value=3$\sigma$ and min\_npix=120 as the astrodendro parameters, we identified 270 starless cores and 224 bound starless cores.
For comparison, we show the CMF derived with the above parameters in Figure \ref{fig:cmf} (b).
The turnover masses of CMFs for starless cores and bound starless cores are $\sim$0.07 $M_\odot$ and
$\sim$0.17 $M_\odot$, respectively.
The differences in turnover masses of each CMF for different parameters are within 1 mass bin.
Thus, we conclude that the dendrogram's parameters do not change the turnover mass dramatically.

We calculated completeness by inserting into the map artificial cores that have a size corresponding to the beam size and FWHM line width of 3 channels (0.3 km/s). The total fluxes are calculated by assuming optically thin emission of C$^{18}$O ($J$=1--0) with $T_\mathrm{ex}$ and $X_\mathrm{C^{18}O}$ derived in Figure \ref{fig:column_density} (a) and a central mass of each mass bin.
We inserted one core to the data which position is random in trunks with avoiding the center overlaps observed cores.
Then, we applied astrodendro to check if it is identified as a {\it leaf}. By repeating the procedure 1000 times for each mass bin, we calculated the detection probability. The 90 \% completeness limits are shown as vertical dashed lines in Figure \ref{fig:cmf}.

\citet{ikeda09} derived
the CMF in almost the same region using the same line C$^{18}$O ($J$=1--0).
They derived the turnover mass at $\sim$ 5 $M_\odot$, about 20 times larger than our value. They
suggested that their turnover mass is an artifact of the poor angular resolution.
The effect of angular resolution on the turnover mass is also discussed in \citet{reid10}.
There are two main differences between \citet{ikeda09}'s and our analyses. One is the core identification method adopted. \citet{ikeda09} used clumpfind algorithm \citep{williams94} which tends to define a core as a structure larger than that identified with dendrogram since clumpfind algorithm allocates all pixels above a threshold to one of the cores. The more important difference is the angular resolution.
In fact, applying the dendrogram to the NRO 45-m only data with 26\arcsec \ resolution, \citet{takemura20} derived the turnover mass of about 0.5 M$_\odot$. This is about 5 times larger than that derived in this study.
If the turnover mass depends only on the angular resolution, the artificial turnover mass obtained from the CARMA-NRO data 
is
about 0.5/$(26/8)^2$ M$_\odot$ $\sim$ 0.05 M$_\odot$. On the other hand, our obtained turnover mass for the bound cores is about 0.1 M$_\odot$, larger than the value expected from the difference of the angular resolution (0.05 M$_\odot$).
Thus, we believe that we constrain the true turnover mass for the bound cores reasonably well.

\section{Discussion}

The IMF in the ONC region is reproduced from our derived CMFs if it is assumed that (1) the star formation efficiency (SFE) of individual cores is constant over the whole mass range as discussed by \citet{alves07} and (2) the SFE of individual cores is 100\%. However, assuming a SFE of 100\% is unphysical, because mass-loss through a protostellar jet is a necessary part of the accretion process, with theoretical models, simulations and observations suggesting that $\sim 30$\% of the accreting mass is lost that way \citep[see the review by][]{pudritz07}. Furthermore, the feedback from outflows can also disperse part of the core mass, with the combined effect of jets and outflow feedback leading to a SFE of order 30\% \citep[e.g.,][]{federrath14}. Thus, our results suggest that mass accretion onto the cores from a larger reservoir must be an ongoing process.

According to the standard scenario of star formation, the
prestellar cores must be self-gravitating to initiate star formation.
Assuming that all the self-gravitating starless cores ($\alpha_{\rm vir} < 2$) form stars within a few free-fall times, we can evaluate the future star formation rate in this region.
Assuming that the star formation timescale is about three times the free-fall time with the mean density of bound starless cores of 4$\times$10$^4$ cm$^{-3}$,
the future star formation rate is calculated to be 1$\times$10$^{-4}$ $M_\odot$ yr$^{-1}$.
This is almost comparable to
the recent star formation rate obtained in Section \ref{subsec:YSO}.
Thus, our results seem to suggest that
self-gravitating cores are likely to be direct progenitors of stars in the ONC region. However, we can not rule out the possibility of star formation from the gravitationally-unbound cores since the star formation rate would be only doubled even if all the starless cores form stars within a few free-fall time.
Recent studies suggest that the majority of the cores are unbound \citep{maruta10,kirk17}, and such cores become gravitationally-bound or disperse eventually \citep{chen20,smullen20}.

If the accretion plays a role in determining the final stellar mass, we expect that our identified protostellar core population has a larger mean mass compared to that of the starless cores.
The mean masses of starless cores and protostellar cores are $\sim$0.19 $M_\odot$ and $\sim$0.67 $M_\odot$, respectively.
For protostellar core masses, we do not include the masses of protostars located inside. This larger mean mass for the protostellar cores is consistent with the idea that the starless cores gain significant gas from the surroundings during star formation.
The importance of the mass accretion onto the cores is also pointed out by \citet{dib10}.

\section{Conclusions}
In this Letter we have compared for the first time the CMF and IMF in the same region, which is located in the Orion Nebula Cluster. Determinations of the two functions with comparable sensitivities have revealed that the CMF has a turnover mass of $\sim$0.1 $M_\odot$, which is comparable to that of the IMF
\citep[see also][for $\rho$ Oph]{bontemps01}.
This seems to contradict the previous conclusion of a larger turnover mass of CMFs \citep[e.g.][]{nutter07,anathpindika11}.
This difference may simply come from the difference in the angular resolutions of the observations.

To keep the slope of the CMF unchanged over time (so that the resultant IMF resembles the stellar IMF observed), the mass accretion rate onto individual cores should be proportional to $M_{\rm core}$ if the timescale of the accretion is more or less constant.
The importance of the mass accretion appears to favor the competitive accretion scenario. However, according to the Bondi-Hoyle-Littleton accretion scenario
the mass accretion rate is proportional to $M_{\rm core}^2$, and thus,
the slopes of the CMFs can change with time \citep[e.g.,][]{goodwin08}.
Numerical simulations also indicate that the actual accretion rates are significantly influenced by environments such as the global gravitational potential, and vary with time \citep[e.g.,][]{klessen01,girichidis12}.
Our result indicates that the mass functions of the stellar seeds already resemble the IMF if the stellar seeds form by the gravitational collapse of the identified prestellar cores.
Recent numerical simulations have pointed out the importance of mass accretion in the evolution of dense cores \citep{haugbolle18,padoan14}, leading to the inertial-inflow scenario.
In contrast to both the core-collapse and the competitive-accretion models, the inertial-inflow model stresses the role of inertial turbulent flows in assembling the stellar mass from a large-scale mass reservoir \citep{padoan20}, even if the CMF and the IMF are very similar \citep{pelkonen20}, as found in this work.

Recent observations have detected the infall motions toward the prestellar cores
\citep{contreras18}.
A significant amount of parent core mass is likely to be blown out by the stellar feedback \citep[e.g.][]{machida12}.
If the stellar feedback is important in determining the core mass, prestellar cores need to gain much more mass from the surroundings \citep{sanhueza19}.

\acknowledgments

Data analysis was carried out on the Multi-wavelength Data Analysis System operated by the Astronomy Data Center (ADC), National Astronomical Observatory of Japan.
Part of this research was carried out at the Jet
Propulsion Laboratory, California Institute of Technology, under a contract
with the National Aeronautics and Space Administration.
P.S. was partially supported by a Grant-in-Aid for Scientific Research (KAKENHI Number 18H01259) of Japan Society for the Promotion of Science (JSPS).
R.S.K.\ acknowledges financial support from the DFG via the collaborative research center (SFB 881, Project-ID 138713538) “The Milky Way System” (subprojects A1, B1, B2, and B8). He also thanks for subsidies from the Heidelberg Cluster of Excellence {\em STRUCTURES} in the framework of Germany’s Excellence Strategy (grant EXC-2181/1 - 390900948) and for funding from the European Research Council (ERC) via the ERC Synergy Grant {\em ECOGAL} (grant 855130).
P.P. acknowledges support by the Spanish MINECO under project AYA2017-88754-P, and financial support from the State Agency for Research of the Spanish Ministry of Science and Innovation through the “Unit of Excellence María de Maeztu 2020-2023” award to the Institute of Cosmos Sciences (CEX2019-000918-M).
V.O., A.S.M.\ and P.S.\ were supported by the Collaborative Research Centre 956, sub-projects C1, A6 and C3, funded by the Deutsche Forschungsgemeinschaft (DFG), project ID 184018867.
We thank the anonymous referee for many useful comments that have improved the presentation.

\facility{CARMA, No:45m, Herschel}
\software{The data analysis in this paper uses python packages Astropy (Astropy Collaboration et al. 2013), SciPy (Jones et al. 2001), Numpy (van der Walt et al. 2011), APLpy (Robitaille \& Bressert 2012), Matplotlib (Hunter 2007) and Astrodendro \citep{rosolowsky08}.}

\clearpage

\bibliography{nakamura}{}
\bibliographystyle{aasjournal}

\end{document}